%% file: ShortPaperMain.tex
\documentclass[manuscript,screen]{acmart}
\usepackage{subcaption}
\usepackage{tipa}
\usepackage{xcolor}

\AtBeginDocument{%
  }

\copyrightyear{2025}
\acmYear{2025}
\setcopyright{rightsretained}
\acmConference[CUI '25]{Proceedings of the 7th ACM Conference on Conversational User Interfaces}{July 8--10, 2025}{Waterloo, ON, Canada}
\acmBooktitle{Proceedings of the 7th ACM Conference on Conversational User Interfaces (CUI '25), July 8--10, 2025, Waterloo, ON, Canada}
\acmDOI{10.1145/3719160.3737625}
\acmISBN{979-8-4007-1527-3/2025/07}

\acmSubmissionID{3391}
\begin{document}

\title{Talking to...uh...um...Machines: The Impact of Disfluent Speech Agents on Partner Models and Perspective Taking}

\author{Rhys Jacka}
\email{rhys.jacka@ucdconnect.ie}
\affiliation{%
  \institution{UCD, School of Information Communication Studies}
  \city{Dublin}
  \country{Republic of Ireland}
}

\author{Paola R. Peña}
\authornotemark[1]
\email{paola.pena@ucdconnect.ie}
\affiliation{
  \institution{UCD, School of Information Communication Studies}
  \city{Dublin}
  \country{Republic of Ireland}
}

\author{Sophie Leonard}
\authornotemark[1]
\email{sophie.leonard@ucd.ie}
\affiliation{
  \institution{UCD, School of Information Communication Studies}
  \city{Dublin}
  \country{Republic of Ireland}
}

\author{Éva Székely}
\authornotemark[1]
\email{szekely@kth.se}
\affiliation{
  \institution{KTH Royal Institute of Technology,
Division of Speech, Music and Hearing}
  \city{Stockholm}
  \country{Sweden}
}

\author{Benjamin R. Cowan}
\authornotemark[1]
\email{benjamin.cowan@ucd.ie}
\affiliation{
  \institution{UCD, School of Information Communication Studies}
  \city{Dublin}
  \country{Republic of Ireland}
}

\begin{abstract}
Speech disfluencies play a role in perspective-taking and audience design in human-human communication (HHC), but little is known about their impact in human-machine dialogue (HMD). In an online Namer-Matcher task, sixty-one participants interacted with a speech agent using either fluent or disfluent speech. Participants completed a partner-modelling questionnaire (PMQ) both before and after the task. Post-interaction evaluations indicated that participants perceived the disfluent agent as more competent, despite no significant differences in pre-task ratings. However, no notable differences were observed in assessments of conversational flexibility or human-likeness. Our findings also reveal evidence of egocentric and allocentric language production when participants interact with speech agents. Interaction with disfluent speech agents appears to increase egocentric communication in comparison to fluent agents. Although the wide credibility intervals mean this effect is not clear-cut. We discuss potential interpretations of this finding, focusing on how disfluencies may impact partner models and language production in HMD.
\end{abstract}



\maketitle

\section{Introduction}
Similar to human-human dialogue (HHD), work in human-machine dialogue (HMD) suggests that audience design also drives our dialogue interactions with conversational user interfaces (CUIs) \cite{rothwell_comparison_2021, branigan_role_2011}, although users can also behave egocentrically in such interactions, using language that is less felicitous for their partner \cite{dombi_common_2022, peña_audience_2023}. Recent work has shown that speech design features, such as accent, can influence our language production with computer partners \cite{cowan_whats_2019}. This supports the notion that design plays a role in audience design in HMD, potentially through impacting users' perceptions of system dialogue capability, termed partner models \cite{doyle_what_2021}. These models refer to users' mental representations of a dialogue partner's communicative competence, including assessments of dependability, human-likeness, and communicative flexibility. These models are shaped by prior expectations and ongoing experience, which is essential for gaining an insight into user expectations and interaction behaviour with a speech agent \cite{doyle_what_2021}.
Our work contributes by further examining the role of speech design, specifically the inclusion of utterance-level linguistic markers common in spoken language performance, such as disfluencies, in HMD. We aim to better understand how the inclusion of disfluencies in automated speech impacts users' perceptions of system dialogue capability and user perspective-taking in utterance production. Using the online Namer-Matcher Task \cite{peña_audience_2023}, participants took turns naming objects and selecting objects based on descriptions from a speech-based conversational agent that produced either fluent or disfluent descriptions. Partner Modelling Questionnaire (PMQ) scores \cite{doyle_partner_2025} showed that while participants initially rated both agents similarly in competence and dependability judgements, post-interaction ratings were significantly lower for the fluent agent compared to the disfluent one. Bayesian mixed effects analysis revealed that, consistent with \cite{peña_audience_2023}, participants more frequently used scalar modifiers (such as “small” to disambiguate) when a larger competitor was present, whether in common ground (shared view) or privileged ground (visible only to the participant), compared to when there was one target (no competitor was present). This indicates that when generating referential expressions in HMD interaction, there is interference from both a user's privileged knowledge and sensitivity to shared visual context. However, the influence of agent fluency remains uncertain. Participants tended to use more scalar modifiers in the disfluent condition when describing target objects in the privileged ground, suggesting greater egocentric language use, while modifier use in common ground remained similar across conditions. The wide credibility intervals indicate considerable variability, suggesting the effect is more complex than a straightforward effect of fluency alone. 

\section{Related Work}
\subsection{Perspective taking and audience design in human-machine dialogue}
Dialogue is a collaborative joint activity \cite{clark_arenas_1992}, within which interlocutors seek to adjust the language they produce to support mutual understanding and coordination of meaning - a process termed audience design \cite{bell_language_1984}, which reflects allocentric communication, where speakers tailor their utterances based on their partner's perspective or knowledge. This activity relies on perspective-taking, informed by assumptions of shared knowledge (termed common ground- \citet{clark_using_1996, horton_when_1996}), partners' communicative competence \cite{loy_perspective_2023, schober_spatial_1993}, contextual demands \cite{hawkins_division_2021, pena2023human}, and a partner's cognitive state \cite{keysar_taking_2000}. Similar audience or recipient design \cite{an_recipient_2021} effects are consistently evident in HMD, whereby people adapt the language they use when speaking to a computer dialogue partner (e.g. \citet{jaber_cooking_2024}; \citet{porcheron_voice_2018}) to ensure communication success \cite{branigan_role_2011}. This adaptation is thought to be informed by people's assumed shared knowledge between themselves and the agent, as well as their partner model- perceptions of a machine partner's dialogue capabilities \cite{doyle_what_2021}.
Speech interface design, specifically aspects such as humanness of the voice \cite{peña_in_prep, luger_like_2016, cowan_what_2017} and accent \cite{cowan_whats_2019}, have been proposed to influence people's partner models, with corresponding changes observed in the use of lexical alternatives \cite{cowan_whats_2019} or levels of informativity \cite{peña_in_prep}. These design-driven shifts in perceived partner capabilities may shape how users engage in audience design (where speakers tailor their utterances based on their partner's knowledge and capabilities), which is commonly seen as the main driver of language production in HMD \cite{rothwell_comparison_2021}. Recent work has emphasised that, similar to HHD \cite{keysar_communication_2007}, users may also exhibit egocentric forms of language production. In such cases, speakers rely more heavily on their own perspective or knowledge, often producing utterances that reflect their private view without adjusting for the listener's access to information. These strategies are not mutually exclusive: users may simultaneously modify their speech to accommodate perceived knowledge of a speech agent \cite{cowan_whats_2019}, while also producing speech taking their own knowledge into account (e.g. \cite{cowan_whats_2019, peña_audience_2023}) during interaction. 
These effects in HMD tend to be studied using referential communication tasks that manipulate the shared knowledge state between interlocutors \cite{peña_audience_2023}, varying what objects appear in common ground (i.e. objects visible to both interlocutors) and privileged ground (i.e. objects visible only to the user \cite{horton_when_1996, peña_audience_2023}. A third context, termed a \textit{one-target} condition is also included, where a target object is present, with no competitors in shared knowledge. This baseline condition helps isolate the influence of shared or private visual information on language production strategies \cite{keysar_taking_2000, peña_audience_2023}. 
In addition to being shaped by the agent's assigned role \cite{peña_audience_2023}, the extent to which users adopt allocentric or egocentric language strategies may also depend on how salient the agent's communicative capabilities appear during interaction. When users perceive a speech agent as less capable or struggling, they may engage in more effortful perspective-taking to accommodate their partner's presumed difficulties\cite{loy_perspective_2023}.
\phantom{Alongside being impacted by agent role \cite{peña_audience_2023}, allocentric or egocentric production may be governed by the salience of a speech agent's capabilities during interaction, whereby people may be more inclined to conduct more perspective-taking effort if the agent is seen as less capable or struggling during dialogue.}

\subsection{Disfluencies in human-machine dialogue}
Within HHD, disfluencies - a spoken language performance behaviour such as hesitation markers (uh, um), repetitions, and self-corrections - serve a communicative function. Disfluencies can signal processing difficulty on behalf of the speaker and help regulate conversational flow within dialogue \cite{clark_using_2002, pickering_toward_2004}. Research shows that speakers use disfluencies more frequently when introducing complex ideas or addressing listeners with limited shared knowledge \cite{brennan_feeling_1995, arnold_if_2007}. Hesitation markers help guide listener comprehension by structuring information clearly \cite{barr_role_2010}, signalling upcoming difficulty, such as unfamiliar or hard-to-name concepts \cite{van_craeyenest_filled_2025}, or indicating a topic shift \cite{pistono_eye-movements_2021}. Furthermore, speakers use hesitation markers to predict upcoming dialogue \cite{arnold_old_2004} or as a stalling strategy \cite{pistono_eye-movements_2021}, allowing a pause for thought \cite{clark_using_2002}, preventing potential interruptions. Consequently, disfluencies have been suggested as a tool for collaborative meaning-making, playing a role in perspective-taking, where speakers adapt their language based on listener needs \cite{arnold_old_2004, brennan_feeling_1995}. Conversational agents that mimic human-like traits and communication behaviours may enable people to interact with them in ways similar to HHD. Research has shown that users adapt their communication styles to the behaviours of voice assistants \cite{koleva_influence_2024, horstmann_communication_2024}, highlighting the need to consider how these stylistic cues influence user engagement and overall satisfaction. Thus, integrating hesitation markers in computer speech may help simulate aspects of HHD by signalling uncertainty, adapting their language and elaborating on their lexical choices. Work has explored this integration, but to improve interaction quality, suggesting that artificial disfluencies can increase user engagement, enhance perceived naturalness, and manage turn-taking \cite{betz_interactive_2018, bohus_managing_2014, chen_effects_2022}.

\subsection{Research Aims \& Hypotheses}
Speech design has been shown to influence user language production and disfluencies can serve as cues that signal processing difficulty to dialogue partners. Thus, using disfluency in machine speech could possibly increase the salience of speech agent capability. In this case, disfluencies may make users more sensitive to the automated partner's capabilities, perspective, and knowledge state, leading to a more effective interaction. Such research allows us to further explore how utterance design can influence user perceptions and perspective-taking processes in HMD, making a theoretical contribution to the field.  
We hypothesise that:
\begin{itemize}
  \item[H1] There will be a statistically significant effect of disfluencies on partner modelling, whereby the disfluent condition will lead to significant differences in partner modelling scores from baseline levels compared to the fluent condition.
  \item[H2] Scalar modifier use will vary across perspective conditions: participants will use modifiers more in the privileged and common ground conditions compared to the one target condition, with higher usage expected in common ground. 
  \item[H3] There will be an interaction effect between perspective conditions and agent disfluency. Specifically, we predict that participants will be more sensitive to their partner's knowledge state when interacting with a disfluent agent, resulting in reduced use of scalar modifiers in the privileged ground condition, compared to interactions with a fluent agent.  
\end{itemize}

\section{Methodology}
\subsection{Participants}
Sixty-one native English-speaking participants from Ireland were recruited via Prolific (mean age = 36, SD = 10.81; 21 male, 34 female, 1 gender queer) and took part in the study, after 23 exclusions for inattention (N=8, e.g. switching screens >= 5 times), eligibility (N=11), or insufficient data (N=4, e.g. skipped rounds). Participants were randomly assigned to the Fluent (N = 30) or Disfluent (N = 26) condition. All had normal or corrected vision and hearing, with no reported speech or cognitive impairments. The study received low-risk ethics approval, and participants were compensated €6.

\subsection{Design and Materials}
\subsubsection{Namer-Matcher Task:}
\begin{figure}[ht]
    \centering
    \begin{subfigure}[b]{0.45\linewidth}
        \centering
        \includegraphics[width=0.5\linewidth]{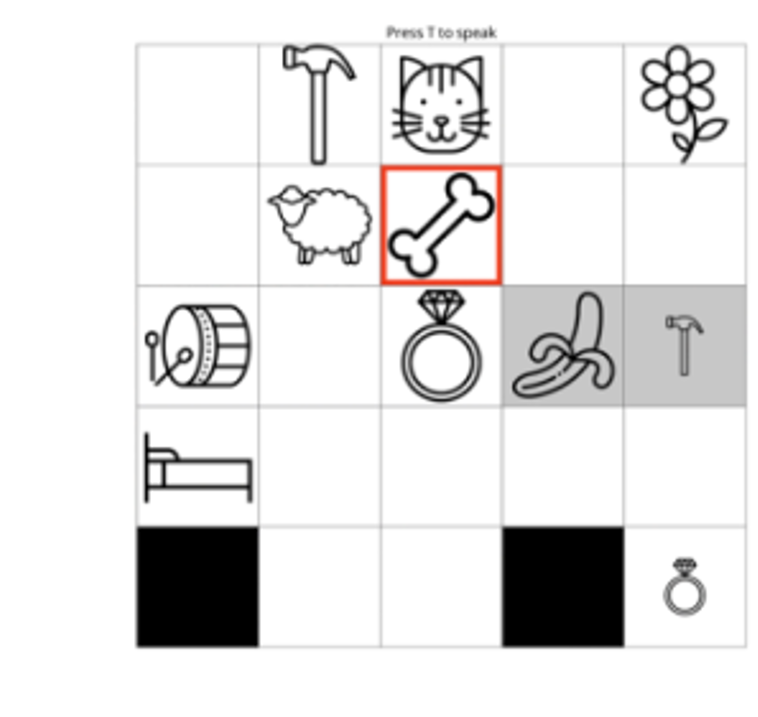}
        \caption{One Target Condition: The target had no competitors.}
        \label{fig:1a}
        \Description{A 5x5 grid containing various black-and-white clipart images. One bone image, highlighted with a red border, appears in second row, third column (the target). No other matching images are present in the grid, indicating there are no competitor objects. This represents the One Target Condition.}
    \end{subfigure}
    \begin{subfigure}[b]{0.45\linewidth}
        \centering
        \includegraphics[width=0.5\linewidth]{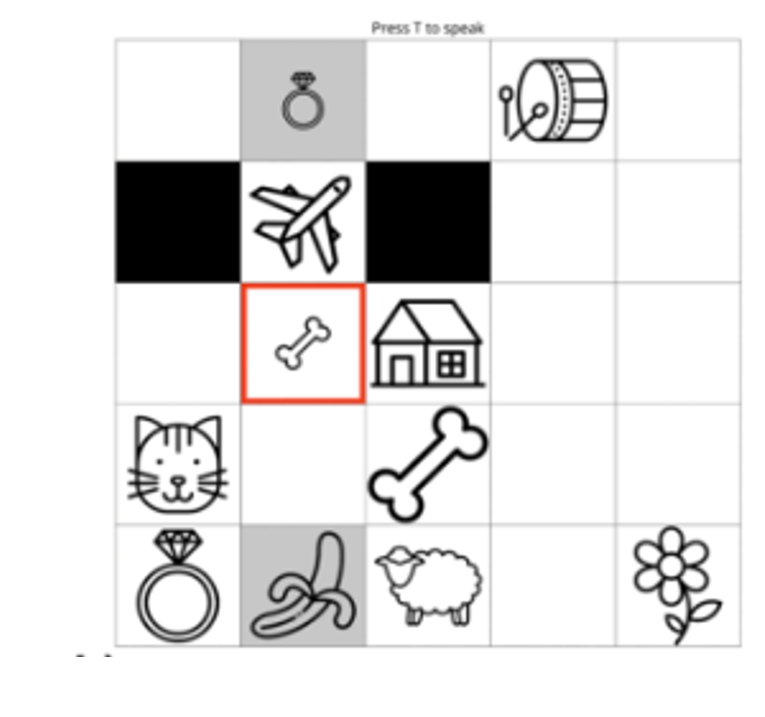}
        \caption{Common Ground Condition: The target object has a competitor visible to both.}
        \label{fig:1b}
        \Description{A 5x5 grid with black-and-white clipart images. The grid shows two identical bone images: one in the third row, second column outlined in red (the target), and another in the fourth row, third column. Both bone images are in visible cells, indicating that the competitor is visible to both the speaker and the listener. This represents the Common Ground condition.}
    \end{subfigure}
        \begin{subfigure}[b]{0.45\linewidth}
        \centering
        \includegraphics[width=0.5\linewidth]{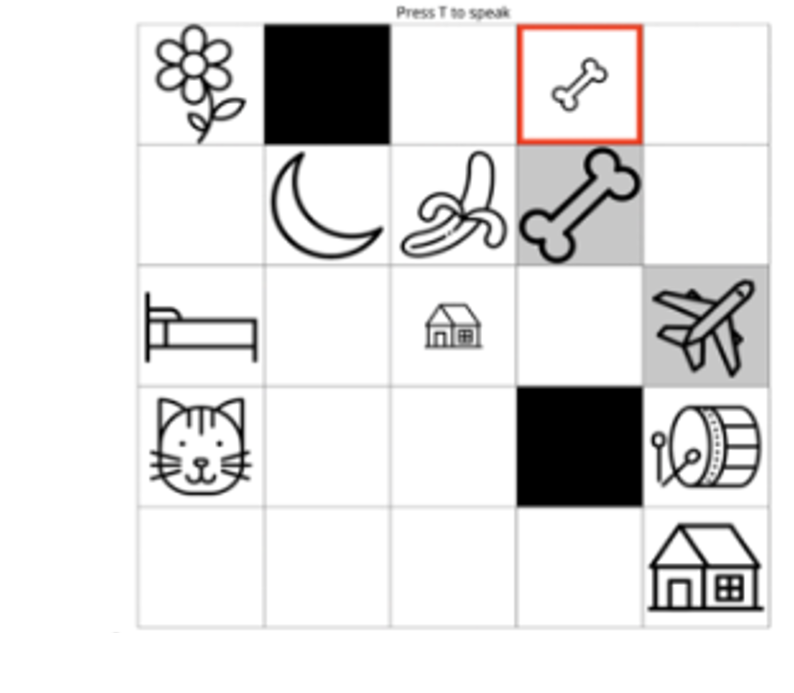}
        \caption{Privileged Ground Condition: The target object has a competitor visible only to the participant.}
        \label{fig:1c}
        \Description{A 5x5 grid displaying various black-and-white clipart images. The grid contains two identical bone images: one in the first row, fourth column, highlighted with a red border (the target), and another in the second row, fourth column, which is shaded to indicate it is visible only to the participant. The condition is labeled as "Privileged Ground," where the competitor object is visible only to the participant.}
    \end{subfigure}
    \caption{Image replicated from Peña et al. (2023) showing experiment perspective conditions.}
    \label{fig:1}
\end{figure}

The study adapted the online Namer-Matcher Task used in \citet{peña_audience_2023}, in which participants alternated between naming and matching turns on a 5×5 grid. In naming turns, they described a target object (highlighted with a red box); in matching turns, they selected an object based on their partner's description. Crucially, visual perspective was manipulated: objects on white backgrounds were shared (common ground - See Fig. - ~\ref{fig:1b}), while those on grey backgrounds were visible only to the participant (privileged ground - See Fig. ~\ref{fig:1c}). Black squares represented the partner's exclusive view.

The game featured monochrome line drawings of common objects. Six items (bone, ring, hammer, car, moon and house) served as experimental targets; seven others were fillers. Each grid included 11 objects (nine in common ground, two in privileged ground) with object placement randomised, but the target order was pre-determined to avoid repetition. The presentation order of the grids and the designation of target objects were also predetermined. The task comprised 30 matching-naming turns (6 practice, 24 experimental).

\subsubsection{Perspective Conditions}
Participants named targets across three perspective conditions: (1) no competitor (One target condition - See Fig. ~\ref{fig:1a}), (2) larger competitor in common ground visible to both (See Fig.~\ref{fig:1b}), and (3) larger competitor in privileged ground visible only to the participant (See Fig.~\ref{fig:1c}). Each condition included six filler trials (participants named objects without competitors) to obscure the study's aims. Smaller objects were always the target, and conditions were pseudo-randomised to prevent consecutive repetitions.

\subsubsection{Speech Agent Conditions}
Participants interacted with a simulated speech agent using pre-recorded Text-To-Speech (TTS) audio based on spontaneous Hiberno-English male speech \cite{ferstl_investigating_2018, miniota_hi_2023}. Using a reverse Wizard of Oz method \cite{branigan_role_2011}, the agent named images fluently or disfluently, with initial hesitation markers (“uh, um”). To minimise linguistic priming, the agent only named filler objects during matching turns, and participants received no feedback on their selections.

\subsection{Dependent Variables - Measures}
\subsubsection{Partner Modelling Questionnaire (PMQ)}
To assess participant's partner models, the PMQ \cite{doyle_partner_2025} was administered before and after the task, assessing perceptions of the agent's competence and dependability (9 items,  Cronbach \textipa{A}: pre= .91; post= .93),  human-likeness (6 items, Cronbach \textipa{A}: pre= .87;  post= .88) and communicative flexibility (3 items,  Cronbach \textipa{A}: pre= .53, post= .76).
\subsubsection{Scalar Modifier Use}
Scalar modifier use (“small” or synonyms) was coded as a binary variable “1” if used appropriately with the target noun (e.g., “small bone”), 0 otherwise.  Utterances were coded as NA if they included a scalar modifier without the target noun; provided directional information (e.g., N= “The small ring, bottom left hand corner”), or self-corrected their description (e.g., N= “The ring - the small ring”). Identified the target using another object as reference (e.g., N= “Click the hammer, beside the cat”).

\section{Procedure}
Following \citet{peña_audience_2023}, participants were recruited and pre-screened on Prolific for Hiberno-English fluency with functional microphones and speakers. After providing informed consent, they completed a short demographic questionnaire and confirmed normal or corrected vision and hearing. Ineligible participants were excluded from the analysis. Participants first completed the PMQ \cite{doyle_partner_2025}, and then received instructions for the Namer-Matcher task, which explained their alternating roles as Namer (describing a target image) and Matcher (selecting the image based on their partner's description). Instructions highlighted the difference in visual perspective—some items would be visible only to them (grey squares) or only to the agent (black squares), and a visual guide was provided. Once participants clicked the start button, the system simulated a connection process, displaying a 30-second “Finding partner…” loading screen before introducing the speech agent partner. This framing was done to support the illusion of a real-time partner and reinforce the impression that the agent was an independent conversational partner, which can impact how people evaluate computer interaction \cite{nass_are_1999, peña_audience_2023}. Participants were randomly assigned to either a fluent or disfluent agent condition and completed six practice trials followed by 24 experimental trials. Afterwards, they completed the PMQ again, a brief post-task questionnaire, and were then debriefed and compensated.

\section{Results \& Discussion}
\subsection{Analysis Approach} 
We analysed scalar modifier use using Bayesian generalised linear mixed effects modelling (GLMM) with Bernoulli distribution and logit link, implemented via the ‘brms' package in R version 4.3.2. Speech Agent (Fluent vs. Disfluent) and Perspective(One Target vs. Privileged Ground vs. Common Ground) conditions were included as fixed effects in the model, while participant and trial stimulus were included as random intercepts. We selected a Bayesian approach to allow for flexibility in estimating effects where prior research has suggested near-zero values for scalar modifier use (namely for the one target condition in this study) and to incorporate this prior knowledge directly into model estimation. Priors were informed by \citet{peña_audience_2023}, using their reported estimates and standard errors for scalar modifier use in the Privileged Ground (B = 7.88, SE = 1.15) and Common Ground (B = 9.96, SE = 1.19) conditions relative to the Target condition to define normal priors. A weakly informative normal prior was used for the intercept to allow moderate baseline variation without unduly influencing the model. An exponential prior was applied to group-level standard deviations to constrain extreme variability. All predictors were sum-coded. Posterior distributions were estimated using four chains of 4000 iterations (1000 warm-up), and all models converged (all $\hat{R} = 1.00$, effective sample sizes > 2000).
In regards to the PMQ, data from the pre- and post-questionnaires were scored according to \citet{doyle_partner_2025}. 2 (Time: Pre vs. Post) x2 (Speech Agent: Disfluent vs. Fluent) Mixed ANOVAs with Bonferroni corrections were used to assess reported differences in the three PMQ subscales: Communicative Competence and Dependability, Human-Likeness and Communicative Flexibility.

\subsection{Results}
\subsubsection{Partner Modelling Questionnaire}
\begin{figure}[ht]
    \centering
    \begin{subfigure}[b]{0.45\linewidth}
        \centering
        \includegraphics[width=1\linewidth]{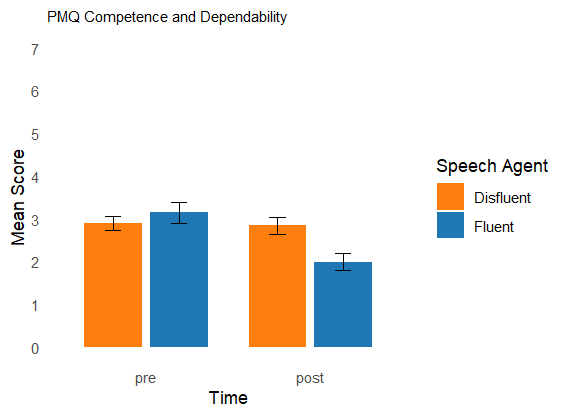}
        \caption{Competence \& Dependability}
        \label{fig:2a}
        \Description{Bar chart titled "PMQ Competence and Dependability" showing mean scores on the y-axis (ranging from 0 to 7) and time points "pre" and "post" on the x-axis. Two bars are shown for each time point: one for the disfluent speech agent (orange) and one for the fluent agent (blue). At the pre-test, both agents received similar mean ratings around 3–3.5. At the post-test, the disfluent agent was rated higher than the fluent agent, which showed a notable drop in mean score. Error bars represent standard error.}
    \end{subfigure}
    \begin{subfigure}[b]{0.45\linewidth}
        \centering
        \includegraphics[width=1\linewidth]{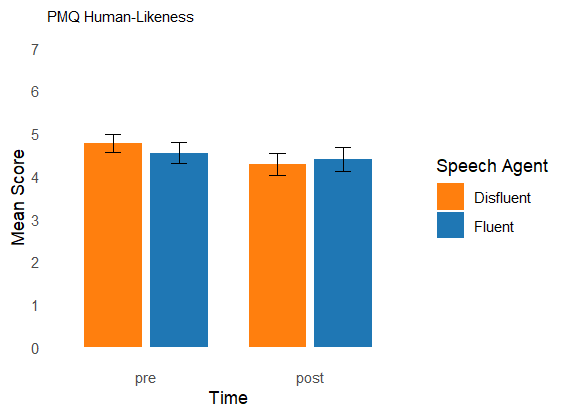}
        \caption{Human Likeness}
        \label{fig:2b}
        \Description{Bar chart titled "PMQ Human-Likeness" showing mean scores (0 to 7) on the y-axis and time points "pre" and "post" on the x-axis. Each time point includes two bars: one for the disfluent speech agent (orange) and one for the fluent agent (blue). At the pre-test, both agents were rated similarly, with mean scores near 4.5. At the post-test, both agents again received similar ratings around 4, with minimal change. Error bars represent standard error.}
    \end{subfigure}
        \begin{subfigure}[b]{0.45\linewidth}
        \centering
        \includegraphics[width=1\linewidth]{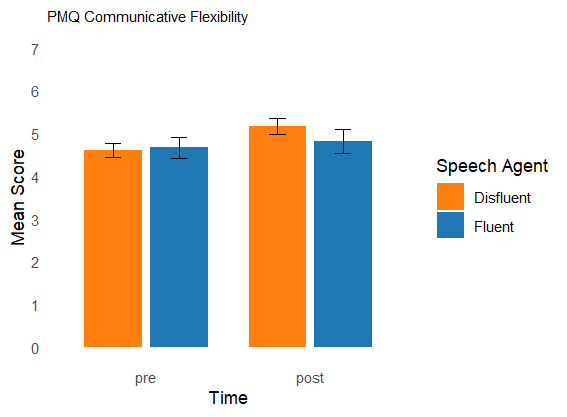}
        \caption{Communicative Flexibility}
        \label{fig:2c}
        \Description{Bar chart titled "PMQ Communicative Flexibility" with mean scores (0 to 7) on the y-axis and time points "pre" and "post" on the x-axis. Two bars are shown for each time point: one for the disfluent speech agent (orange) and one for the fluent agent (blue). At the pre-test, both agents were rated similarly just below a score of 4.5. At the post-test, both ratings increased slightly, with the disfluent agent rated marginally higher at just above 5 than the fluent agent. Error bars indicate standard error.}
    \end{subfigure}
    \caption{Bar plots portraying changes in PMQ score for each of the three subscales. (a) Competence and Dependability, (b) Human-Likeness and (c ) Communicative Flexibility, before (pre) and after (post) engaging with the perspective-taking task for both fluency conditions, N=56.}
    \label{fig:2}
\end{figure}

Three 2 (Time: Pre vs. Post- within participants) × 2 (Speech Agent: Disfluent vs. Fluent between participants) mixed ANOVAs were conducted on PMQ subscales\footnote{A Bonferroni adjusted alpha level of .0167 was used to control for multiple comparisons}. For the Competence and Dependability subscale (See Fig.~\ref{fig:2a}), a statistically significant main effect of Time [F(1, 54) = 16.22, p < .001, partial $\eta^2$ = .072] and a significant Time by Speech Agent interaction was observed [F(1, 54) = 13.77, p < .001, partial $\eta^2$ = .062]. Post hoc comparisons revealed a significant decrease from pre to post-task in the Fluent condition [t(54) = 5.68, p < .001], but not the Disfluent condition [t(54) = 0.22, p = .830]. No other statistically significant effects were observed for the Communicative Flexibility or the Human Likeness subscales, which gives support for H1.

\subsubsection{Scalar Modifier Use in Perspective-taking Task}
\begin{figure}[htbp]
  \centering
  \includegraphics[width=0.7\linewidth]{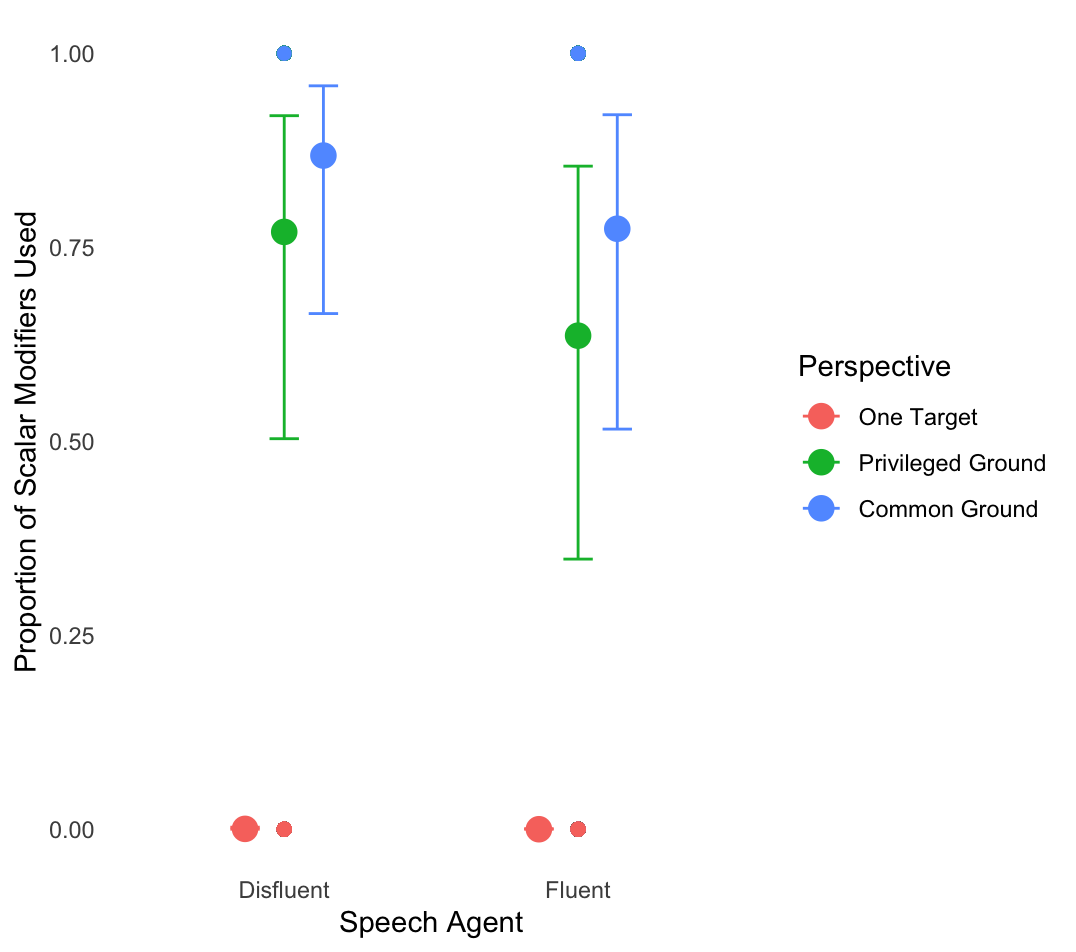} 
  \caption{Mean and standard error of proportion of scalar modifiers used between partner conditions across perspective conditions, N= 56.}
  \label{fig:3}
  \Description{Dot-and-error bar plot showing the proportion of scalar modifiers used by participants, grouped by Speech Agent (Disfluent vs. Fluent) on the x-axis. The y-axis represents the proportion of scalar modifiers used (ranging from 0.00 to 1.00). Data points are colour-coded by Perspective condition: red for One Target, green for Privileged Ground, and blue for Common Ground. In both agent conditions, the Common Ground condition shows the highest proportion of scalar modifier use, followed by Privileged Ground, with the One Target condition near zero. Error bars indicate variability across participants.}
\end{figure}

There was strong evidence for the main effects of Perspective conditions on scalar modifier use. Participants were more likely to produce scalar modifiers in the Privileged Ground condition (b = 8.89, SE = 0.75, 95\% CrI [7.44, 10.38]) and the Common Ground condition (b = 9.58, SE = 0.76, 95\% CrI [8.10, 11.09]) compared to the One Target condition. Meanwhile, a strong difference was not observed between the Common Ground and Privileged Ground conditions (b = 0.68, SE = 0.35, 95\% CrI [0.02, 1.39]), suggesting that participants increased their use of scalar modifiers similarly for stimuli in both conditions. The effect of Speech Agent fluency on scalar modifier use remained uncertain, as the main effect of fluency was negative but not credibly different from zero (b = -6.35, SE = 5.28, 95\% CrI [-19.39, 0.84]). Similarly, the interaction effects between Speech Agent and Perspective conditions displayed wide credibility intervals. A strong difference was not observed between Common Ground and Privileged Ground for either the fluent (b = 0.68, SE = 0.35, 95\% CrI [0.02, 1.39]) or the disfluent condition (b = 0.68, SE = 0.32, 95\% CrI [0.06, 1.32]). This indicates that while the overall level of scalar modifier use was higher for both Common and Privileged Ground compared to One Target, participants were slightly more likely to use scalar modifiers in the Common Ground condition compared to the Privileged Ground condition, and this effect was consistent across both fluent and disfluent speech (See Fig.~\ref{fig:3}).

\section{Discussion}
Our analysis revealed that before interaction, participants initially rated both the fluent and disfluent agents as similarly competent and dependable. Yet after interaction, the disfluent agent was deemed significantly more competent and dependable. This supports previous work surrounding the dynamic nature of partner models over interaction \cite{gessinger_chatgpt_2025}, whilst suggesting that people may see the fluent partner as a less competent dialogue partner. Findings align with \citet{peña_audience_2023}, whereby participants were more likely to use scalar modifiers in both the Common and Privileged Ground conditions compared to the One Target condition, indicating that they adapted their language production in response to the perspective conditions - exhibiting both egocentric behaviour (using scalar modifiers when competitors were only seen by the participant) and allocentric behaviour (using scalar modifiers more frequently to disambiguate competitors in common ground). The impact of fluency on this perspective effect is more ambiguous, with speakers seemingly experiencing more interference from privileged ground when interacting with the disfluent as opposed to fluent speech agent. These findings are interpreted below. 

\subsection{Partner modelling dynamics and the impact of disfluency}
Our results demonstrate that people tended to rate the disfluent agent as more competent and dependable post-interaction, compared to a fluent agent. This finding is surprising given our expectation that a disfluent agent would be perceived as less competent than a standard fluent agent, considering that disfluencies are markers for speech processing difficulties and uncertainty \cite{clark_using_2002, pickering_toward_2004, van_craeyenest_filled_2025}. This might be due to people perceiving the disfluent agent as being more reflective of the game state, thus more true to the nature of the application. Although, participants were informed that the task involved ambiguous information states. A mismatch in expectations can negatively affect perceptions of virtual agents \cite{glikson_human_2020, rheu_systematic_2021}, with context and task proposed as playing a key role in setting such expectations \cite{luger_like_2016}, which may explain the fluent condition post-interaction. Therefore, people may have felt that the disfluent agent was more authentic to the task state and rated its competence accordingly. Our work also supports the notion that partner models are dynamic, changing from initial expectations \cite{gessinger_chatgpt_2025}. Thus, it seems that disfluent agents positively impact partner models, particularly competence judgments, from a user's initial global model of speech agents measured before interaction. While these findings point to useful design insights, they also raise important ethical considerations. The subtle use of disfluencies to influence users' perceptions of agent competence could be problematic in trust-sensitive settings such as healthcare, education, or finance. In these contexts, speech cues that artificially influence perceived understanding or credibility may undermine transparency and user autonomy \cite{glikson_human_2020, rheu_systematic_2021}. Future work, should explore how to balance the benefits of naturalistic design in speech agents alongside ethical responsibilities around user trust, informed decision-making, and system transparency \cite{nass_are_1999, cowan_what_2017}. Similar to previous work on the longitudinal development of partner models with systems like ChatGPT \cite{gessinger_chatgpt_2025}, future research could explore how perceptions evolve over time and how repeated interactions with disfluent agents shape people's global partner models of speech agents \cite{galati_attenuating_2010, gessinger_chatgpt_2025}. Work could also look to further explore how the reflection of ambiguity within contexts and tasks within speech agent design influences partner models, to further support the findings of this work.

\subsection{Deciphering the effect of disfluencies on perspective taking}
As highlighted, participants who interacted with the disfluent agent appeared to behave more egocentrically when forming their utternances, producing referential expressions that reflected their own visual perspective, with less adjustment for what the agent could or could not see. This contrasts with audience design, in which speakers deliberately tailor utterances to their partner's presumed knowledge. The increased use of scalar modifiers in the privileged ground condition might suggest egocentrism, as participants were considering their own knowledge rather than the agent's perspective. However, these findings should be interpreted with caution due to wide credibility intervals (see Fig.~\ref{fig:3}), and the motivations behind this behaviour could be interpreted in several ways. An interpretation that emerged from our partner model findings was that people are more egocentric when engaging with a disfluent speech agent because they perceive the agent to be more capable of interpreting overly informative utterances. This was because in the PMQ findings participants rated the disfluent agent to be more competent and more capable compared to the fluent agent. Enabling the participants to generate descriptions that are more relevant to their perspective, reducing their perspective-taking effort \cite{loy_perspective_2023, peña_audience_2023}. Alternatively, increased scalar modifier use might reflect an audience design strategy. Participants may have been encoding contrastive or distinguishing features to support interpretation for the disfluent speech agent - a dialogue partner possibly perceived as experiencing communicative difficulty. Aligning with use of salience markers \cite{flavell_young_1981}, where speakers explicitly encode and produce salient features in images to support their partner in identifying the right object, especially in contexts involving ambiguity or knowledge asymmetries \cite{pena2023human}. Rather than being egocentric, such behaviour reflects an audience design strategy in which speakers tailor their utterances based on the presumed perspective or needs of their dialogue partner \cite{kraljic_prosodic_2005}. Although disfluent speech agents were generally perceived as more competent and dependable post-interaction, the agents disfluency may have signalled processing difficulty during dialogue, prompting participants to increase referential specificity to support understanding \cite{arnold_if_2007, brennan_partner-specific_2009}. Thus, participants may have adjusted their language to accommodate both the referential context and their partner's perceived cognitive state. Future work should further explore these interpretations and investigate if there is a balance between the two perspectives in HMD. This could involve studies that examine user behaviour over extended interactions, or incorporate post-task qualitative feedback measures to better understand how users' partner models evolve, and how their language strategies may shift over time.

\subsection{Limitations \& Future Directions}
While this study provides valuable insights into how speech disfluency and referential perspective influence language production in human–machine dialogue, it does have its contextual constraints. Much like \citet{peña_audience_2023} experimental dialogue partners, our agent was simulated using pre-recorded speech, while effective for control this likely differs from naturalistic use of real-time conversational agents, where agents dynamically respond. A corresponding limitation, also noted in previous work, is that participants' interactions with the speech agents occurred over a relatively short duration, which limits our ability to examine how partner models or communicative strategies might develop over time. This short duration, coupled with the absence of a qualitative follow-up, also restricts our understanding of participants' underlying motivations, whether their utterances were driven by egocentric planning or deliberate audience design. Additionally, the study focused exclusively on one form of disfluency, initial hesitation markers, which may not fully capture the range of effects that more naturalistic and varied disfluency patterns could have on user perception, linguistic adaptations and partner models. These limitations point toward potential future directions from this research and elaborate further, including studies in more naturalistic contexts, and incorporating qualitative user reflections.

\section{Conclusion}
This study investigated how referential visibility and agent fluency affect speakers' use of scalar modifiers in human-machine dialogue. The results showed a clear effect of visibility: participants used more scalar modifiers in both the Common Ground and Privileged Ground conditions compared to the One Target condition, suggesting they adapted their lexical choices to account for potential uncertainty in the visual context. It is cautiously noted that participants also seemed to increase their use of scalar modifiers when interacting with a disfluent agent, particularly in the Privileged Ground condition, even though the agent could not see the competitor object. This may indicate an audience design strategy based on the presumed perspective and needs of the agent. However, the motivations behind these adaptations remain nuanced: while increased modifier use in the Privileged Condition may reflect egocentric planning based on private knowledge, it may also indicate an audience design response to the disfluent agent's perceived communicative challenges. That said, this effect was uncertain, as statistical models showed high variability and wide credibility intervals. These results highlight that referential visibility strongly influences linguistic adaptation in HMD, while the impact of disfluency remains ambiguous and may depend on how users interpret and engage with the agent.

\begin{acks}
This work was conducted with the financial support of the Research Ireland Centre for Research Training in Digitally-Enhanced Reality (d-real) under Grant No. 18/CRT/6224. Additional support for this research was provided by Research Ireland at ADAPT, the Research Ireland Centre for AI-Driven Digital Content Technology at University College Dublin [13/RC/2106\_P2].    
\end{acks}

\bibliographystyle{ACM-Reference-Format}
\input{ShortPaperMain.bbl}

\end{document}